# An Overview of Multiple Sequence Alignment Systems

.      Fahad Saeed and Ashfaq Khokhar

We care about the sequence alignments in the computational biology because it gives biologists useful information about different aspects. For example, it can tell us about the evolution of the organisms, we can see which regions of a gene (or its derived protein) are susceptible to mutation and which can have one residue replaced by another without changing function, we can study Homologous genes and can uncover paralogs and Orthologs genes that are evolutionary related.

In problems such as the construction of an evolutionary tree based on sequence data, or in protein engineering, where a multiple alignment of related sequences may often yield the most helpful information on the design of a new protein, a molecular biologist must compare more than two sequences simultaneously. [1]

A multiple sequence alignment (MSA) arranges protein sequences into a rectangular array with the goal that residues in a given column are homologous (derived from a single position in an ancestral sequence), superposable (in a rigid local structural alignment) or play a common functional role. Although these three criteria are essentially equivalent for closely related proteins, sequence, structure and function diverge over evolutionary time and different criteria may result in different alignments. Manually refined alignments continue to be superior to purely automated methods; there is therefore a continuous effort to improve the biological accuracy of MSA tools. Additionally, the high computational cost of most naive algorithms motivates improvements in speed and memory usage to accommodate the rapid increase in available sequence data. [2]

## 2. Exact Algorithms

A straightforward dynamic programming algorithm in the k-dimensional edit graph formed from k strings solves the Multiple Alignment problem. For example, suppose that we have three sequences **u**, **v**, and **w**, and that we want to find the "best" alignment of all three. Every multiple alignment of three sequences corresponds to a path in the three-dimensional Manhattan like edit graph. In this case, one can apply the same logic as we did for two dimensions to arrive at a dynamic programming recurrence, this time with more terms to consider. To get to vertex (i, j, k) in a three-dimensional edit graph, you could come from any of the following predecessors (note that δ(x, y, z) denotes the score of a column with letters x, y, and z, as in the figure shown below.

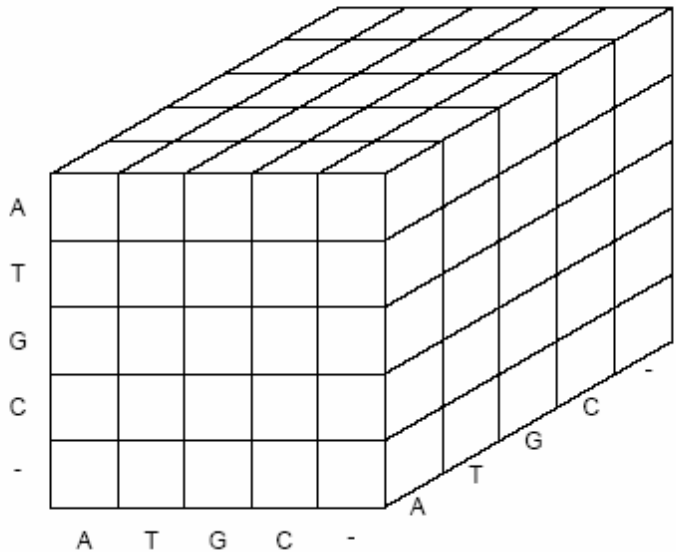

The scoring matrix, δ, used in a three-sequence alignment.

1. $(i-1, j, k)$ for score $\delta(u_i,-,-)$
2. $(i, j-1, k)$ for score $\delta(-, v_j, -)$
3. $(i, j, k-1)$ for score $\delta(-,-, w_k)$
4. $(i-1, j-1, k)$ for score $\delta(u_i, v_j, -)$
5. $(i-1, j, k-1)$ for score $\delta(u_i, -, w_k)$
6. $(i, j-1, k-1)$ for score $\delta(-, v_j, w_k)$
7. $(i-1, j-1, k-1)$ for score $\delta(u_i, v_j, w_k)$

We create a three-dimensional dynamic programming array **s** and it is easy to see that the recurrence for $s_{i,j,k}$ in the three-dimensional case is similar to the recurrence in the two-dimensional case.

$s_{i,j,k}$ = max {
   $s_{i-1,j,k} + \delta(v_i, -, -)$
   $s_{i,j-1,k} + \delta(-, w_j, -)$
   $s_{i,j,k-1} + \delta(-, -, u_k)$
   $s_{i-1,j-1,k} + \delta(v_i, w_j, -)$
   $s_{i-1,j,k-1} + \delta(v_i, -, u_k)$
   $s_{i,j-1,k-1} + \delta(-, w_j, u_k)$
   $s_{i-1,j-1,k-1} + \delta(v_i, w_j, u_k)$
}

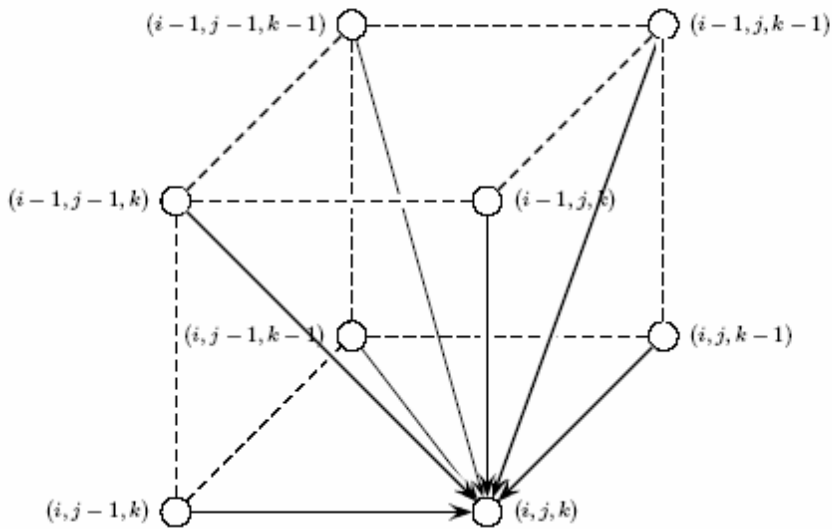

A cell in the alignment graph between three sequences.

Unfortunately, in the case of k sequences, the running time of this approach is $O((2n)k)$, so some improvements of the exact algorithm, and many heuristics for suboptimal multiple alignments, have been proposed. A good heuristic would be to compute all optimal pairwise alignments between every pair of strings and then combine them together in such a way that pairwise alignments induced by the multiple alignment are close to the optimal ones. Unfortunately, it is not always possible to combine optimal pairwise alignments into a multiple alignment since some pair-wise alignments may be Incompatible.

As can be seen that the methods of dynamic programming that give best possible alignment are not practically feasible. The alignments that are useful are in the order of hundred's of sequences and these methods may not be feasible for more than 7 sequences. Therefore methods are required that would allow to get alignments for more sequences.

One of the widely used methods is Iterative Algorithms. Mostly these algorithms are used in conjunction with other methods. Iterative algorithms are a kind of tune up algorithms that are used so that the alignment that is achieved is more 'accurate'.

## 3. Progressive Alignment

### 3.1 Introduction to Progressive Alignment

One method of performing a heuristic alignment search is the progressive technique that builds up a final MSA by first performing a series of pairwise alignments on successively less closely related sequences. The most commonly used heuristic methods are based on the progressive-alignment strategy. Such methods begin by aligning the two most closely related sequences first and then successively aligning the next most closely related sequence in the query set to the alignment produced in the previous step. Although the performance depends on the quality of the initial alignment especially and it degrades significantly when all of the sequences in the set are rather distantly related, progressive alignment methods are efficient enough to implement on a large scale for many sequences.

A very popular progressive alignment method is the Clustal [8] family, especially the weighted variant ClustalW [9] which could be assecced by many web portals like [GenomeNet](#) [10], [EBI](#) [11], and [EMBNet](#) [12]. Another common progressive alignment method called T-Coffee [13] is slower than Clustal and its derivatives but generally produces more accurate alignments for distantly related sequence sets. T-coffee uses the output from Clustal as well as another local alignment program LALIGN, which finds multiple regions of local alignment between two sequences. But the progressive methods are not guaranteed to converge to a global optimum, alignment quality can be difficult to evaluate and their true biological significance can be obscure.

### 3.2 T-Coffee

### 3.2.1 Introduction to T-Coffee

Before start talking about T-Coffee, we first have a glimpse of ClustalW. That will help us to understand why T-Coffee is proposed. ClustalW alignment algorithm consists of three steps: 1) Alignment scores are used to build a distance matrix by taking the divergence of the sequence into account. 2) A guide (phylogenetic) tree is created from the distance matrix using the Neighbor-Joining method. This guide tree has branches of different lengths, and their length is proportional to the estimated divergence along each branch. 3) Progressive alignment of the sequence is done by following the branch order of the guide tree. The alignment of the sequences is guided by the phylogenetic relationship indicated by the tree. Although ClustalW is the most widely used implementation and successful in a wide variety of cases, this method suffers from its greediness. Errors made in the first alignments cannot be rectified later as the rest of the sequences are added in. T-Coffee is an attempt to minimize that effect, and although T-Coffee itself is also a greedy progressive method, it allows for much better use of information in the early stages. T-Coffee (Tree-based Consistency Objective Function for alignment Evaluation) has two main features. First, it makes good use of heterogeneous data sources and the data from these sources are provided to T-Coffee via a mixture of local and global pair-wise alignments. Second, it provides a simple and flexible and, most importantly, accurate solution to the problem of how to combine information of this sort. The overall performance on 141 test case alignments from the BaliBase collection proves its efficiency and accuracy. T-Coffee is a progressive alignment with an ability to consider information from all of the sequences during each alignment step, not just those being aligned at that stage.

### 3.2.2 T-Coffee Algorithm
**Step 1. Generating a Primary Library of Alignments**

**Two Data Sources for Input Library**
The primary library contains a set of pair-wise alignments between all of the sequences to be aligned. The alignments are not required to be consistent. E.g. there can be two or more different alignments of the same pair of sequences. Two alignment sources for each pair of sequences, one local and one global, are used in T-coffee. The global alignments are constructed using ClustalW on the sequences, two at a time. This is used to give one full-length alignment between each pair of sequences. The local alignments are the ten top scoring non-intersecting local alignments, between each pair of sequences, gathered using the Lalign program of the FASTA package with default parameters. In the library, each alignment is represented as a list of pair-wise residue matches (e.g. residue x of sequence A is aligned with residue y of sequence B).

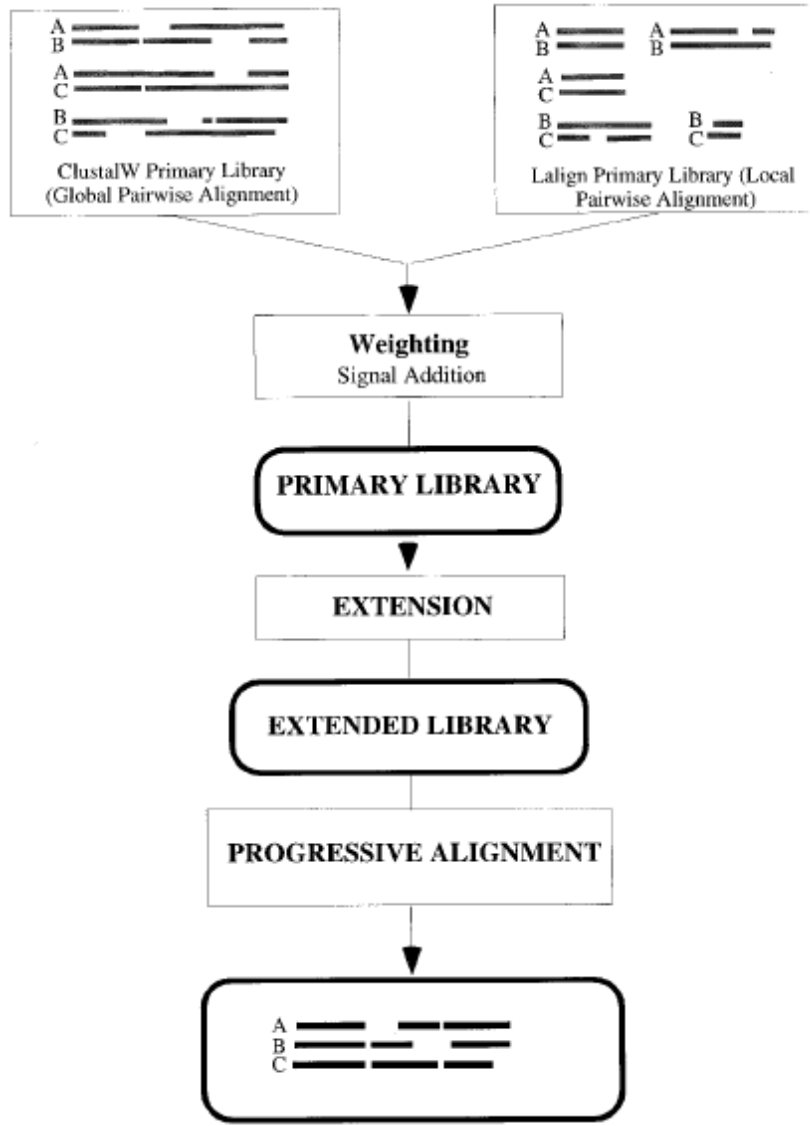

Figure 1. Main Steps in T-Coffee.
Square blocks designate procedures while rounded blocks indicate data

**Compute the Primary Library Weights**

T-Coffee assigns a weight to each pair of aligned residues in the library in order to reflect the correctness of an aligned residue pair. T-coffee considers each of these pairs as a constraint. All of these constraints are not equally important. The parts of alignments, which are more likely to be correct, are taken into account when computing the multiple alignment and the priority is given to the most reliable residue pairs by using a weighting scheme.

**Combination of the Libraries**
Our aim is the efficient combination of local and global alignment information. This is achieved by pooling the ClustalW and Lalign primary libraries. If any pair is duplicated between the two libraries, it is merged into a single entry that has a weight equal to the sum of the two weights. Otherwise, a new entry is created for the pair being considered. Pairs of residues that did not occur are not represented; they will be considered to have a weight of zero by default.

**Step 2. Extending the library**
The overall idea is to combine information in such a manner that the final weight, for any pair of residues, reflects some of the information contained in the whole library. It is based on taking each aligned residue pair from the library and checking the alignment of the two residues with residues from the remaining sequences. First, we carry out the extension on each pair of residues in a pair of sequences. The initial weight for a pair of residues in two sequences can be set to the primary weight. And the final weight associated with that pair will be the sum of all the weights gathered through the examination of all the triplets involving that pair. The more intermediate sequences supporting the alignment of that pair, the higher its weight. Once the extension on the pair of sequences is complete, this procedure is repeated for each remaining pair of sequences.

The worst case complexity for this step is $O(N^3L^2)$, where N is the number of sequences and L is the average sequence length.

    2.1) Carry out the extension on one pair of residues. The extension needs to align one aligned residue pair through the remaining (N-2) sequences. Step 2.1) needs $O(N)$.

    2.2) Now look at one pair of sequences. There are at most (L*L) pairs of residues for one sequence pair. The worst case for this step is $O(L*L)$. This will only occur when all the included pair-wise alignments are totally inconsistent. In practice, step 2.2 needs $O(L)$.

    2.3) The extension is repeated for each pair of sequences. There are totally N(N-1)/2 pairs of sequences.

Therefore, the time complexity for step 2 is $O(N^3L^2)$, and in practice, the complexity is closer to $O(N^3L)$.

**Step 3 - Progressive Alignment**
Pair-wise alignments are first made to produce a distance matrix between all the sequences, which in turn is used to produce a guide tree using the neighbor-joining method.

The closest two sequences on the tree are aligned first using dynamic programming. This alignment uses the weights in the extended library to align the residues in the two sequences. Then the next closest two sequences are aligned or a sequence is added to the existing alignment of the first two sequences, depending which is suggested by the guide tree.
This continues until all the sequences have been aligned.

**Time Complexity for T-Coffee**
The complexity of the whole procedure is $O(N^2L^2) + O(N^3L) + O(N^3) + O(NL^2)$ where N is the number of sequences and L is the average sequence length. $O(N^2L^2)$ is associated with the computation of the pair-wise library, $O(N^3L)$ with the extension, $O(N^3)$ with the computation of the NJ tree and $O(NL^2)$ with the computation of the progressive alignment. In fact, because of $L \gg N$, $O(N^2L^2) + O(NL^2) \gg O(N^3L)$, which means the time required for library and the alignment is much larger than the extension.

## 3.2.3 Experiment Result

Table 1. The effect of combining local and global alignments

| Name | Protocol | | | Cat1 (81) | Cat2 (23) | Cat3 (4) | Cat4 (12) | Cat5 (11) | Total (141) | Significance |
|---|---|---|---|---|---|---|---|---|---|---|
| C | ClustalW pw | ... | ... | 70.6 | 26.7 | 43.0 | 56.0 | 60.0 | 58.9 | 7.8ª |
| CE | ClustalW pw | ... | extend | 77.1 | 33.6 | 47.6 | 64.8 | 75.9 | 66.3 | 17.7ª |
| L | ... | Lalign pw | ... | 65.4 | 12.1 | 22.8 | 53.9 | 66.0 | 52.0 | 7.8ª |
| LE | ... | Lalign pw | extend | 72.6 | 25.6 | 47.2 | 77.5 | 85.5 | 64.2 | 16.3ª |
| CL | ClustalW pw | Lalign pw | ... | 76.2 | 32.0 | 48.3 | 76.2 | 74.6 | 66.5 | 12.1ª |
| CLE | ClustalW pw | Lalign pw | extend | **80.7** | **37.3** | **52.9** | **83.2** | **88.7** | **72.1** | ... |

Protocol shows the way the library was created. ClustalW pw and Lalign pw show the pair-wise alignments computed with one of these programs, using default parameters. Extend indicates that the library was extended before progressive alignment. CLE uses a combination of ClustalW and Lalign alignments and library extension. Cat1 to Cat5 are the five reference categories of BaliBase; numbers in parentheses indicate the number of alignments in a category. The average accuracy is then given for each protocol. The best accuracies in each column are shown in bold and underlined. Total gives the average accuracy across all 141 test alignments. The last column shows the percentage of times that CLE is outperformed by each other protocol. The statistical significance of the improvement of CLE over each protocol is shown by a (P < 0.001).

# 4. Iterative Alignment

## 4.1 Introduction
A set of methods to produce MSAs while reducing the errors inherent in progressive methods are classified as "iterative" because they work similarly to progressive methods but repeatedly realign the initial sequences as well as adding new sequences to the growing MSA. One reason progressive methods are so strongly dependent on a high-quality initial alignment is the fact that these alignments are always incorporated into the final result - that is, once a sequence has been aligned into the MSA, its alignment is not considered further. This approximation improves efficiency at the cost of accuracy. By contrast, iterative methods can return to previously calculated pairwise alignments or sub-MSAs incorporating subsets of the query sequence as a means of optimizing a general objective function such as finding a high-quality alignment score.

A variety of subtly different iteration methods have been implemented and made available in software packages; reviews and comparisons have been useful but generally refrain from choosing a "best" technique[3]. The software package PRRN/PRRP uses a hill-climbing algorithm to optimize its MSA alignment score [4] and iteratively corrects both alignment weights and locally divergent or "gappy" regions of the growing MSA[5].PRRP performs best when refining an alignment previously constructed by a faster method. The alignment of individual motifs is then achieved with a matrix representation similar to a dot-matrix plot in a pairwise alignment. An alternative method that uses fast local alignments as anchor points or "seeds" for a slower global-alignment procedure is implemented in the CHAOS/DIALIGN suite[6].

A third popular iteration-based method called MUSCLE (multiple sequence alignment by log-expectation) improves on progressive methods with a more accurate distance measure to assess the relatedness of two sequences. The distance measure is updated between iteration stages (although, in its original form, MUSCLE contained only 2-3 iterations depending on whether refinement was enabled).

Two of the most accurate Multiple Alignment Sequence programs at the moment are MUSCLE and ProbCons [7]. Each of these methods for the multiple Sequence Alignment are discussed below.

## 4.2 MUSCLE

The basic strategy used in MUSCLE is that a progressive alignemnt is built to which a horizontal refinement is applied.

**Algorithm Overview**

*Stage 1: draft progressive*
The first stage builds a progressive alignment.

*Similarity measure*
The similarity of each pair of sequences is computed, either using *k*-mer counting or by constructing a global alignment of the pair and determining the fractional identity.

*Distance estimate*
A triangular distance matrix is computed from the pairwise similarities.

*Tree construction*
A tree is constructed from the distance matrix using UPGMA or neighbor-joining, and a root is identified.

*Progressive alignment*
A progressive alignment is built by following the branching order of the tree, yielding a multiple alignment of all input sequences at the root.

*Stage 2: improved progressive*
The second stage attempts to improve the tree and builds a new progressive alignment according to this tree. This stage may be iterated.

*Similarity measure*
The similarity of each pair of sequences is computed using fractional identity computed from their mutual alignment in the current multiple alignments.

*Tree construction*
A tree is constructed by computing a Kimura distance matrix and applying a clustering method to this matrix.

*Tree comparison*
The previous and new trees are compared, identifying the set of internal nodes for which the branching order has changed. If Stage 2 has executed more than once, and the number of changed nodes has not decreased, the process of improving the tree is considered to have converged and iteration terminates.

*Progressive alignment*
A new progressive alignment is built. The existing alignment is retained of each subtree for which the branching order is unchanged; new alignments are created for the (possibly empty) set of changed nodes. When the alignment at the root is completed, the algorithm may terminate, return to step 2.1 or go to Stage 3.

***Stage 3: refinement***
The third stage performs iterative refinement using a variant of tree-dependent restricted partitioning [12].

*Choice of bipartition*
An edge is deleted from the tree, dividing the sequences into two disjoint subsets (a bipartition). Edges are visiting in order of decreasing distance from the root.

*Profile extraction*
The profile (multiple alignment) of each subset is extracted from the current multiple alignment. Columns containing no residues (i.e., indels only) are discarded.*Re-alignment*
The two profiles obtained in step 3.2 are re-aligned to each other using profile-profile alignment.

*Accept/reject*
The SP score of the multiple alignments implied by the new profile-profile alignment is computed. If the score increases, the new alignment is retained, otherwise it is discarded. If all edges have been visited without a change being retained, or if a user-defined maximum number of iterations have been reached, the algorithm is terminated, otherwise it returns to step 3.1. Visiting edges in order of decreasing distance from the root has the effect of first realigning individual sequences, then closely related groups.

**Algorithm Elements**

In the following we discuss the different aspects of MUSCLE algorithm.

*Objective score*

In its refinement stage, MUSCLE seeks to maximize an objective score, i.e. a function that maps a multiple sequence alignment to a real number which is designed to give larger values to better alignments. MUSCLE uses the *sum-of-pairs* (SP) score, defined to be the sum over pairs of sequences of their alignment scores.

The alignment score of a pair of sequences is computed as the sum of substitution matrix scores for each aligned pair of residues, plus gap penalties.

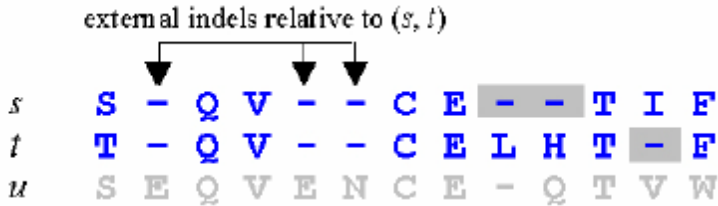

Gap penalties in the SP score

*Progressive alignment*

Progressive alignment requires a rooted binary tree in which each sequence is assigned to leaf. The tree is created by clustering a triangular matrix containing a distance measure for each pair of sequences. At each internal node, profile-profile alignment is used to align the existing alignments of the two child sub trees, and the new alignment is assigned to that node. A multiple alignment of all input sequences is produced at the root node.

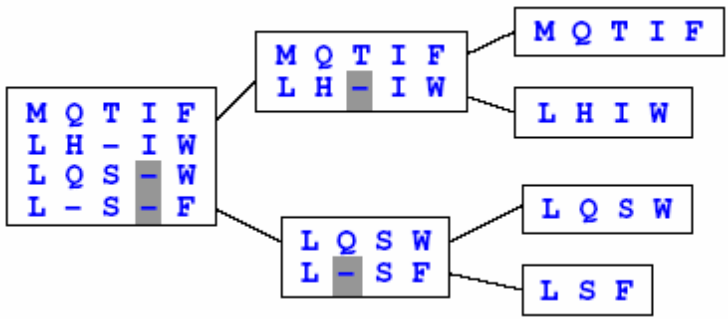

*Similarity measures*

We use the term *similarity* for a measure on a pair of sequences that indicates their degree of evolutionary divergence (the sequences are assumed to be related). MUSCLE uses two types of similarity measure: the fractional identity $D$ computed from a global alignment of the two sequences, and measures obtained by $k$-mer counting.

The similarity measure b/w X and Y is defined as

$$F = \Sigma_\tau \min [n_X(\tau), n_Y(\tau)] / [\min (L_X, L_Y) - k + 1].$$

Where Here $\tau$ is a $k$-mer, $L_X$, $L_Y$ are the sequence lengths, and $n_X(\tau)$ and $n_Y(\tau)$ are the number of times $\tau$ occurs in X and Y respectively.

*Distance measures*

Given a similarity value, we wish to estimate an additive distance measure. An additive measure distance measure d(A, B) between two sequences A and B satisfies d(A, B) = d(A, C) + d(C, B) for any third sequence C, assuming that
A, B and C are all related.

*Tree construction*
Given a distance matrix, a binary tree is constructed by clustering. Two methods are implemented: neighbor-joining and UPGMA.

*Profile functions*
A commonly used profile function is the sequence-weighted sum of substitution matrix scores for each pair of letters, selecting one from each column (PSP, for profile SP):

PSP$xy$ = Σ$i$ Σ$j$ $f$ $xi$ $f$ $yj$ $Sij$.
Note that $Sij$ = log ($pij$ / $pipj$) , so
PSP$xy$ = Σ$i$ Σ$j$ $f$ $xi$ $f$ $yj$ log ($pij$ / $pi$ $pj$).

MUSCLE implements PSP functions based on the 200 PAM matrix of [33] and the 240 PAM VTML matrix [34]. In addition to PSP, MUSCLE implements a function we call the *log-expectation* (LE) score.

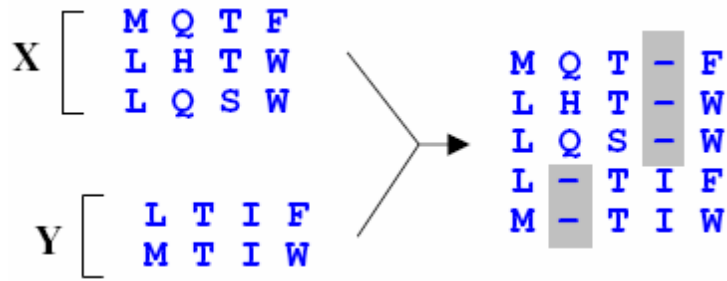

**Figure: Profile-Profile Alignment**

*Tree comparison*
In progressive alignment, two subtrees will produce identical alignments if they have the same set of sequences at their leaves and the same branching orders (topologies). We exploit this observation to optimize the progressive alignment in Stage 2 of MUSCLE, which begins by constructing a new tree. Unchanged subtrees are identified, and their alignments are retained.

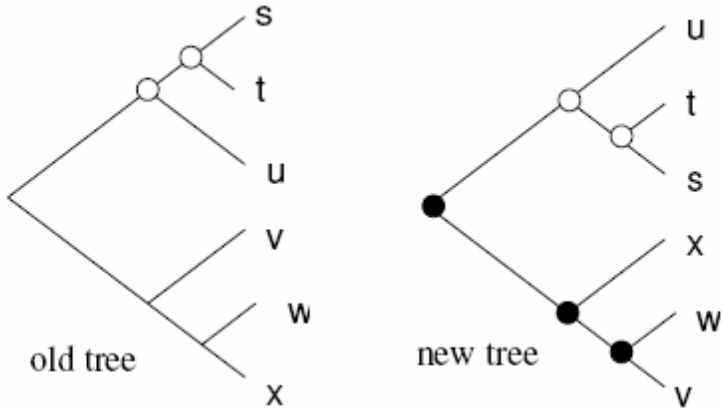

**Tree comparison.** Two trees are compared in order to identify those nodes that have the same branching orders within subtree rotation (white). If a progressive alignment has been created using to the old tree, then alignments at these nodes can be retained as the same result would be produced at those nodes by the new tree. New alignments are needed at the changed (black) nodes only.

## *Complexity of MUSCLE*

The complexity of MUSCLE is summarized in Table below. We assume $LP = O(L + N)$, the e-string construction for the root alignment, and a fixed number of refinement iterations.

Table 2: Complexity of MUSCLE. Here we show the big-O asymptotic complexity of the elements of MUSCLE as a function of $L$, the typical sequence length, and $N$, the number of sequences, retaining the highest-order terms in $N$ with $L$ fixed and vice versa.

| Step | O(Space) | O(Time) |
|---|---|---|
| K-mer distance matrix | $N^2 + L$ | $N^2 L$ |
| UPGMA | $N^2$ | $N^2$ |
| Progressive (one iteration) | $L_p^2 = NL + L^2$ | $L_p^2 = N^2 + L^2$ |
| Progressive (root alignment) | $NL_p = N^2 + NL$ | $NL_p \log N = N^2 \log N + NL \log N$ |
| Progressive (N iterations + root) | $N^2 + NL + L^2$ | $N^3 + NL^2$ |
| Refinement (one edge) | $NL_p + L_p^2 = N^2 + L^2$ | $N^2 L_p + L_p^2 = N^3 + L^2$ |
| Refinement (N edges) | $N^2 + L^2$ | $N^4 + NL^2$ |
| TOTAL | $N^2 + L^2$ | $N^4 + NL^2$ |

Table 3: Accuracy scores. The average accuracy, measured by the $Q$ score, is reported for each method on each set of reference alignments.

| Method | PREFAB | BAliBASE | SABmark | SMART |
|---|---|---|---|---|
| MUSCLE | 0.648 | 0.896 | 0.430 | 0.856 |
| MUSCLE-prog | 0.634 | 0.883 | 0.427 | 0.837 |
| FFTNS1 | 0.619 | 0.844 | 0.376 | 0.815 |
| MUSCLE-fast | 0.616 | 0.849 | 0.396 | 0.813 |
| CLUSTALW | 0.588 | 0.860 | 0.404 | 0.823 |
| POA-blast | 0.577 | 0.839 | 0.352 | 0.788 |
| POA | 0.576 | 0.834 | 0.280 | 0.797 |

MUSCLE demonstrates improvements in accuracy and reductions in computational complexity by exploiting a range of existing and new algorithmic techniques. While the design–typically for practical multiple sequence alignment tools–arguably lacks elegance and theoretical coherence, useful improvements were achieved through a number of factors. Most important of these were selection of heuristics, close attention to details of the implementation, and careful evaluation of the impact of different elements of the algorithm on speed and accuracy. MUSCLE enables high-throughput applications to achieve average accuracy comparable to the most accurate tools previously available, which we expect to be increasingly important in view of the continuing rapid growth in sequence data.

## 5. Hidden Markov Chain based Alignment - Probcons

### 5.1 Motivation for ProbCons

Obtaining accurate alignments is a difficult computational problem because of not only the high computational cost but also the lack of proper objective functions for measuring alignment quality. Probcons introduced the notion of *probabilistic consistency*, a novel scoring function for multiple sequence comparisons. ProbCons is a practical tool for progressive protein multiple sequence alignment based on probabilistic consistency.

Direct application of dynamic programming is too inefficient for alignment of more than a few sequences. Instead, a variety of heuristic strategies have been proposed. By far, the most popular heuristic strategies involve tree-based *progressive alignment* in which groups of sequences are assembled into a complete multiple alignment via several pairwise alignment steps. As with any hierarchical approach, however, errors at early stages in the alignment not only propagate to the final alignment but also may increase the likelihood of misalignment due to incorrect conservation signals. Post-processing steps such as iterative refinement alleviate some of the errors made during progressive alignment.

ProbCons is a pair-hidden Markov model-based progressive alignment algorithm that primarily differs from most typical approaches in its use of *maximum expected accuracy* rather than Viterbi alignment, and of the *probabilistic consistency transformation* to incorporate multiple sequence conservation information during pairwise alignment.

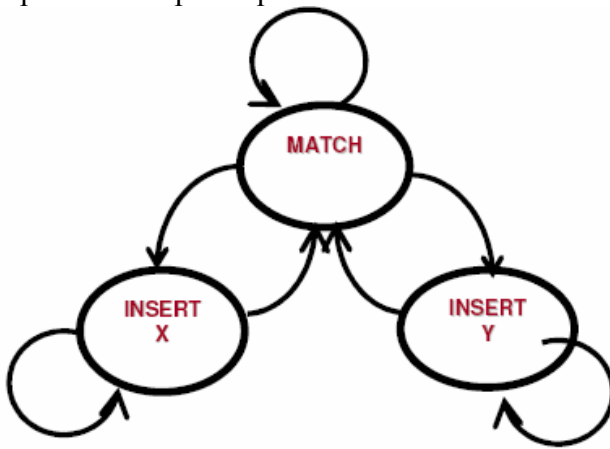

Emission probabilities, which correspond to traditional substitution scores, are based on the BLOSUM62 matrix. Transition probabilities, which correspond to gap penalties, are trained with unsupervised expectation maximization (EM).

*Consistency-based* schemes take the view that "prevention is the best medicine." Note that for any multiple alignment, the induced pairwise alignments are necessarily *consistent*— that is, given a multiple alignment containing three sequence $x$, $y$, and $z$, if position $xi$ aligns with position $zk$ and position $zk$ aligns with $yj$ in the projected $x$–$z$ and $z$–$y$ alignments, then $xi$ must align with $yj$ in the projected $x$–$y$ alignment. Consistency-based techniques apply this principle in reverse, using evidence from intermediate sequences to guide the pairwise alignment of $x$ and $y$, such as needed during the steps of a progressive alignment. By adjusting the score for an $xi \sim yj$ residue pairing according to support from some position $zk$ that aligns to both $xi$ and $yj$ in the respective $x$–$z$ and $y$–$z$ pairwise comparisons, consistency-based objective functions incorporate multiple sequence information in scoring pairwise alignments.

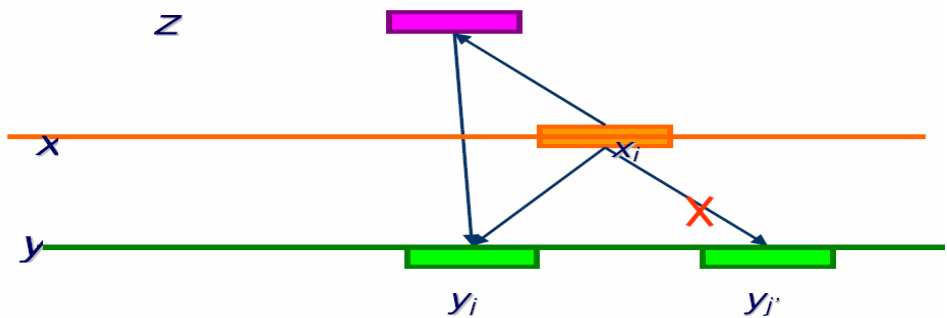

## 5.2 ProbCons algorithm

Given $m$ sequences, $S = \{s(1), \ldots, s(m)\}$:

*Step 1: Computation of posterior-probability matrices*
For every pair of sequences $x, y \in S$ and all $i \in \{1,\ldots, |x|\}, j \in \{1,\ldots,|y|\}$, compute the matrix $P_{xy}$, where $P_{xy}(i, j) = \mathbf{P}(xi \sim yj \in a^* \mid x, y)$ is the probability that letters $xi$ and $yj$ are paired in $a^*$, an alignment of $x$ and $y$ generated by the model.

*Step 2: Computation of expected accuracies*
Define the expected accuracy of a pairwise alignment $a$ between $x$ and $y$ to be the expected number of correctly aligned pairs of letters, divided by the length of the shorter sequence:

$$\mathbf{E}_{a^*}(accuracy(a,a^*)|x,y) = \frac{1}{\min\{|x|,|y|\}} \sum_{x_i - y_j \in a} \mathbf{P}(x_i \sim y_j \in a^*|x,y).$$

For each pair of sequences $x, y \in S$, compute the alignment $a$ that maximizes expected accuracy by dynamic programming, and set
$E(x, y) = \mathbf{E}_{a^*}(accuracy(a, a^*) \mid x, y).$

*Step 3: Probabilistic consistency transformation*
Reestimate the match quality scores $\mathbf{P}(xi \sim yj \in a^* \mid x, y)$ by applying the *probabilistic consistency transformation*, which incorporates similarity of $x$ and $y$ to other sequences from $S$ into the $x$–$y$ pairwise comparison:

$$\mathbf{P}'(x_i \sim y_j \in a^* | x, y) \leftarrow \frac{1}{|S|} \sum_{z \in S} \sum_{z_k} \mathbf{P}(x_i \sim z_k \in a^* | x, z) \mathbf{P}(z_k \sim y_j \in a^* | z, y).$$

In matrix form, the transformation may be written as

$$P'_{xy} \leftarrow \frac{1}{|S|} \sum_{z \in S} P_{xz} P_{zy}.$$

Since most values in the $P_{xz}$ and $P_{zy}$ matrices will be near zero, the transformation is computed efficiently using *sparse* matrix multiplication by ignoring all entries smaller than a threshold _. This step may be repeated as many times as desired.

*Step 4: Computation of guide tree*
Construct a guide tree for $S$ through hierarchical clustering. As a measure of similarity between two sequences $x$ and $y$ use $E(x, y)$ as computed in Step 2. Define the similarity of two clusters by a weighted average of the pairwise similarities between sequences of the clusters.

*Step 5: Progressive alignment*
Align sequence groups hierarchically according to the order specified in the guide tree. Alignments are scored using a sumof-pairs scoring function in which aligned residues are assigned the transformed match quality scores $\mathbf{P}\_(x_i \sim y_j \_ a^* | x, y)$ and gap penalties are set to zero.

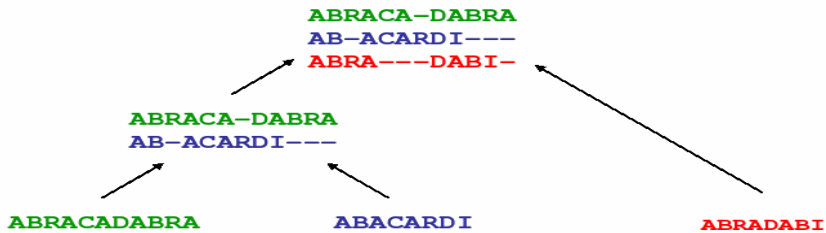

*Post-processing step: Iterative refinement*
Randomly partition alignment into two groups of sequences and realign. This step may be repeated as many times as desired. In this approach, the sequences of the existing multiple alignment are randomly partitioned into two groups of possibly unequal size by randomly assigning each sequence to one of the two groups to be realigned. Subsequently, the same dynamic programming procedure used for progressive alignment is employed to realign the two projected alignments. This refinement procedure can be iterated either for a fixed number of iterations or until convergence; for simplicity, only the former of these options is implemented in ProbCons, where 100 rounds of iterative refinement are applied in the default setting.

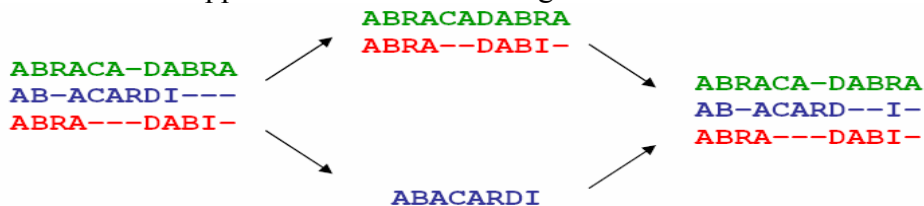

To study gene evolution across a wide range of organisms, biologists need accurate tools for multiple sequence alignment of protein families. Obtaining accurate alignments, however, is a difficult computational problem because of not only the high computational cost but also the lack of proper objective functions for measuring alignment quality. In ProbCons *probabilistic consistency* is introduced, a novel scoring function for multiple sequence comparisons. ProbCons, is a practical tool for progressive protein multiple sequence alignment based on probabilistic consistency, and evaluate its performance on several standard alignment benchmark data sets. On the BAliBASE, SABmark, and PREFAB benchmark alignment databases, ProbCons achieves statistically significant improvement over other leading methods while maintaining practical speed.

## 6. Multiple Sequence Alignment using Genetic Algorithm

### 6.1 Introduction

Genetic algorithms have been used for MSA production in an attempt to broadly simulate the hypothesized evolutionary process that gave rise to the divergence in the query set. The method works by breaking a series of possible MSAs into fragments and repeatedly rearranging those fragments with the introduction of gaps at varying positions. A general objective function is optimized during the simulation, most generally the "sum of pairs" maximization function introduced in dynamic programming-based MSA methods. A technique for protein sequences has been implemented in the software program SAGA (Sequence Alignment by Genetic Algorithm) [14] and its equivalent in RNA is called RAGA [15]. The technique of simulated annealing, by which an existing MSA produced by another method is refined by a series of rearrangements designed to find more optimal regions of alignment space than the one the input alignment already occupies. Like the genetic algorithm method, simulated annealing maximizes an objective function like the sum-of-pairs function. Simulated annealing uses a metaphorical "temperature factor" that determines the rate at which rearrangements proceed and the likelihood of each rearrangement; typical usage alternates periods of high rearrangement rates with relatively low likelihood (to explore more distant regions of alignment space) with periods of lower rates and higher likelihoods to more thoroughly explore local minima near the newly "colonized" regions. This approach has been implemented in the program MSASA (Multiple Sequence Alignment by Simulated Annealing) [16, 17, 18].

### 6.2 Multiple Sequence Alignment using Genetic Algorithm

Genetic Algorithm (GA) is an adaptive method which may be used to solve search and optimization problems. GA is inspired by the mechanism of natural selection where stronger individuals are likely the winners in a competing environment. It is powerful and broadly applicable stochastic search and optimization techniques and is perhaps the most widely known types of evolutionary computation methods today. In general, a GA has five basic components [19]:1) a genetic representation of solution to the problem; 2) a way to create an initial population of solutions; 3) an evaluation function rating solutions in terms of their fitness; 4) genetic operators that alter the genetic composition of children during the reproduction and 5) values for the parameters of genetic algorithms.

Basic idea: Start with a set of solutions called population. Select solutions from one population according to their fitness and use them to form a new population. This is motivated by a hope, that the new population will be better than the old one. The more suitable they are the more chances they have to reproduce.

## 6.2.1 Outline of Genetic Algorithm

i)Choose initial population
ii)Evaluate the fitness of each individual in the population
iii)Repeat
    a)Select best-ranking individuals to reproduce
    b)Breed new generation through genetic operations (crossover and mutation) and give birth to offspring
    c)Evaluate the individual fitness of the offspring
    d)Replace worst ranked part of population with offspring
iv)Until a solution is found that satisfies minimum criteria or a fixed number of generations reached.

## 6.2.2 Apply GA to Multiple Sequence Alignment

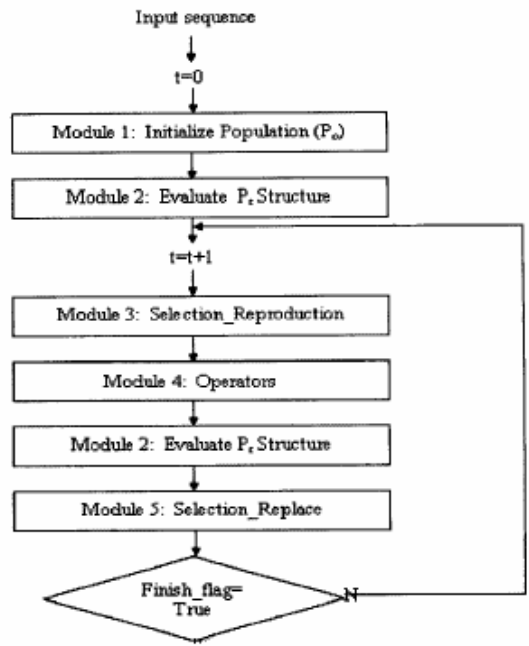

**Module 1- Initialize Population**
Objective: To populate solution randomly.
Description: the population is represented as an array of sequence where each sequence was encoded as an array of character over the alphabet. The symbol "-" will refer to the gap in the alignment which represent an insertion or a deletion of an amino acid residue.

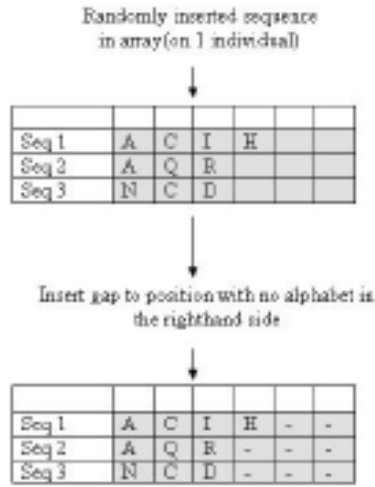

**Module 2 - Evaluate P$_t$ Structure**
Objective: Evaluate and assign a scoring function to each individual in population.
Description: Individual that scores a high fitness function (F) will survive for the next iteration. The Scoring function is as follows:

$$\cos t = \sum_{i=1}^{l}\sigma(S'[i], T'[i])$$

where l=|S'| = |T'|, σ(x,x) = 1 and σ(x,y) = σ(-,y) = σ(x, -) = 0

**Module 3 – Selection Reproduction**
Objective: To select two individuals which have the best fitness function.
Description: the selection probability for each individual is proportional to the fitness function value. In this case, the fitter the individual, the more likely it will be chosen. Compute selection probabilities for the current population based on fitness value. Select two individual randomly based on the selection probabilities to obtain clones which may then be subjected to mutation or recombination.

**Module 4 – Genetic Operators**

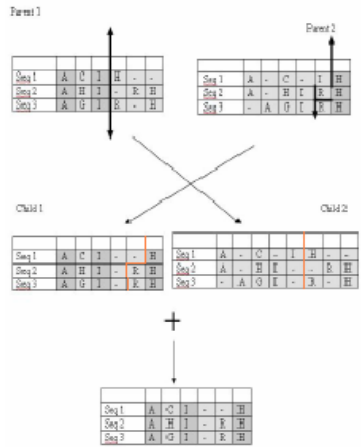

FIGURE 3
ONE POINT CROSSOVER.

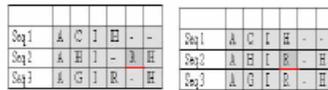

FIGURE 4
2 – OPT MUTATION.

Objective: To explore new region of solution
Description:
Crossover: the crossover operator will use point-to-point crossover where the operator takes two alignment sequences from the population and randomly select a fully matched (no gap) column. After crossover, Child 1 and Child 2 are evaluated. The fittest offspring will survive in the next iteration.
Mutation: The mutation operator picks a random amino acid from a randomly chosen row (sequence) in the alignment and checks whether one of its neighbors has a gap. If this is the case, the algorithms swaps (2-opt) the selected amino acid with a gap neighbor. If both neighbors are gaps, one of them will be picked randomly.

**Module 5 – Selection Replace**
Objective: to replace old individual with new offspring.
Description: this module will replace old individual that have less fitness function and insert new offspring to the population.

### 6.2.3 Experiment Result

The implementation used the protein data sets used by Horg [13], that can be accessed at http://rsdb.csie.ncu.edu.tw/tools/msa.htm. These are protein data sets that were extracted from Swiss-Prot 39.16. Altogether, 8 protein datasets were used. Similarity measures for the pre-alignment and GA are shown in the following table.

| Protein ID | Pre-Alignment | Genetic Algorithm |
|---|---|---|
| P1 | 0.0192 | 0.25 |
| P2 | 0.0020 | 0.56 |
| P3 | 0.6462 | 0.74 |
| P4 | 0.0019 | 0.4 |
| P5 | 0.0028 | 0.5 |
| P6 | 0.0049 | 0.63 |
| P7 | 0.0022 | 0.422 |
| P8 | 0 | 0.32 |

## 7. Multiple Sequence Alignment using Simulated Annealing

### 7.1 Introduction to Simulated Annealing (SA)

Simulated annealing (SA) is a generalization of the Monte Carlo method for examining the equations of state and frozen states of n-body systems [20]. The concept is based on the manner in which liquids freeze or metals recrystallize in the process of annealing. During the annealing state, process will melt and initially it will start at high temperature and it will slowly cool down so that the system will always be at thermodynamic equilibrium. As cooling proceeds, the system becomes more ordered and approaches a "frozen" ground state at T=0. Hence the process can be thought of as an adiabatic approach to the lowest energy state. If the initial temperature of the system is too low or cooling is done insufficiently, slowly the system may become quenched forming defects or freezing out in metal stable states (i.e. trapped in a local minimum energy state). In the original Metropolis scheme was that an initial state of a thermodynamic system was chosen at energy E and temperature T. Holding T constant, the initial configuration is perturbed and the change in energy, dE is computed. If the change in energy is negative the new configuration is accepted. If the change in energy is positive it is accepted with a probability given by the Boltzmann factor exp -(dE/T). These processes are repeated for a number of times to obtain good sampling statistics for the current temperature. The temperature is decremented and the entire process was repeated until a frozen state is achieved at T=0. By analogy the generalization of this Monte Carlo approach to combinatorial problems is straightforward. The current state of the thermodynamic system is analogous to the current solution to the combinatorial problem, the energy equation for the thermodynamic system is analogous to the objective function, and ground state is analogous to the global minimum.

The major advantage of SA over other methods is the ability to avoid becoming trapped at local minima. And the avoidance of entrainment in local minima is dependent on the "annealing schedule", the choice of initial temperature, how many iterations are performed at each temperature, and how much the temperature is decremented at each step as cooling proceeds.

## 7.2 Contact-based Simulated Annealing Protein Sequence Alignment Method [21]

### 7.2.1 The CAO (Contact Accepted mutatiOn) Score Matrix

The CAO substitution matrix is based on an evolutionary Markov model of the protein side-chain contact evolution. The CAO scores can be used to detect both sequence and structure similarities and serve as an intermediate between the sequence-based scores and the structure-based scores. The CAO matrix is composed of 400*400 contact substitution scores, where rows and columns are all the possible contacts of 20 amino acids.

### 7.2.2 Alignment with Simulated Annealing

The following elements must be provided in SA:
1) Solution space
For two sequences with length n and m, the maximum length of an alignment sequence is (n+m).
2) A generator of random changes in solutions
The solution generator should introduce small random changes and allow all possible solutions to be reached.
3) A means of evaluating the problem functions
There are two kinds of scores for accessing the quality of an alignment: the sequence scores and the contact scores.

$$S = \sum_i S_i^{seq} + w(\sum_{contact\ j} S_j^{CAO} + c)$$

- $w$ - the relative weight of CAO scores versus Blosum62 scores.
- $c$ - the matrix constant for CAO scores.
- three parameters: gap-open penalty $p$, gap-extension penalty $q$, and contact-penalty $r$.

4) An annealing schedule - an initial temperature and rules for lowering it as the search progresses.
5) Termination criteria
The algorithm halts when the solution is not changed in three generations searching.

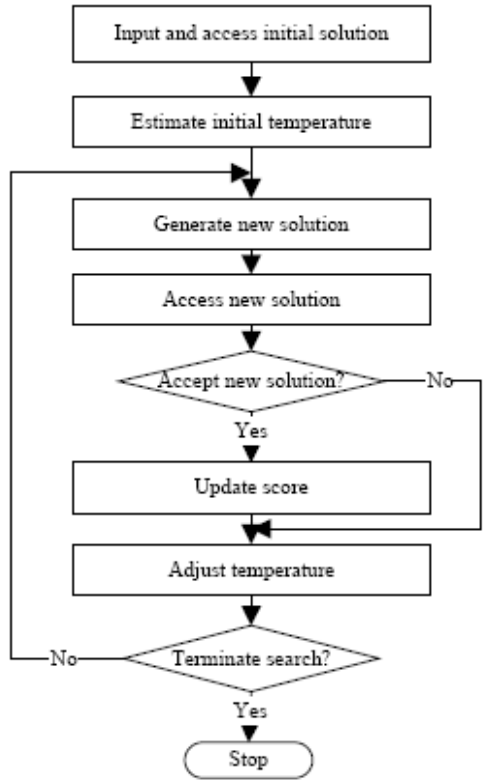

### 7.2.3 Experiment Result

The standard evaluation data is taken from the Homstrad database. Only families containing two single chain sequences were considered (629 alignments in total).

TABLE 1. The optimal parameters of different method

| Method | Matrix | w | c | p | q | r |
|---|---|---|---|---|---|---|
| NW | B | - | - | 11.0 | 2 | - |
| CBDP | B+C | 0.1 | 2.8 | 14.0 | 1.0 | - |
| CBSAA-1 | B+C | 0.2 | 2.9 | 14.0 | 1.0 | - |
| CBSAA-2 | B+C | 0.15 | 3.3 | 15 | 1.0 | 3.0 |

Alignments were performed using CAO (C) or BLOSUM62 (B) matrix
NW: Needleman-Wunsch dynamic programming method
CBDP: contact-based dynamic programming sequence alignment method
CBSAA-1: contact based simulated annealing alignment method without contact penalty
CBSAA-2: contact-based simulated annealing alignment method with contact penalty

TABLE 2. Comparative performances of different alignment methods

| Method | NW | CBDP | CBSAA-1 | CBSAA-2 |
|---|---|---|---|---|
| Accuracy | 78.8% | 81.8% | 84.3% | 85.4% |

## 8. Conclusion

There generally exist three categories of optimization algorithm for multiple alignment; exact, progressive and iterative. Numerous MSA programs have been applied using many techniques and algorithms. Most commonly used techniques are progressive and iterative techniques. The exact method suffers from inexact sequence alignment and lead to an aggressive research on progressive and iterative algorithms.

The exact method can align up to ten closely related sequences. But when the number of sequences becomes larger, the space and time complexity is huge. Progressive alignment constitutes one of the simplest ways to align sequence. This approach has the advantages of speed and simplicity. However the major problem with progressive alignment method is that errors in the initial alignments are the most closely related sequence propagated to the multiple alignment. Iterative alignment methods depend on algorithms that are able to produce an alignment and to refine through a series of iterations until no more improvement can be made. Iterative methods can be deterministic or stochastic, depending on the strategy used to improve the alignment. The simplest iterative strategies are deterministic. It involves extracting the sequence one by one from multiple alignments and realigning them to the remaining sequences. This procedure is terminated when no improvement can be made (convergence). Stochastic iterative methods include Hidden Markov Model (HMM), simulated annealing and evolutionary computation such as genetic algorithms (GAs) and evolutionary programming. Their main advantage is to allow for a good separation between the optimization process and evaluation criteria. It is the objective function that defines the aim of any optimization procedure.

# Reference

1. H. Carillo and D. Lipman, The multiple sequence alignment problem in biology, SIAM Journal of Applied Mathematics, Vol 48, 1988, pp. 1073-1082.
2. Robert C Edgar and Serafim Batzoglou2, Current Opinion in Structural Biology 2006, 16:368–373, DOI 10.1016/j.sbi.2006.04.004
3. 3. Hirosawa M, Totoki Y, Hoshida M, Ishikawa M. (1995). Comprehensive study on iterative algorithms of multiple sequence alignment. *Comput Appl Biosci* 11:13-18
4. Gotoh O. (1996). Significant improvement in accuracy of multiple protein sequence alignments by iterative refinement as assessed by reference to structural alignments. *J Mol Biol* 264(4):823-38.
5. Mount DM. (2004). Bioinformatics: Sequence and Genome Analysis 2nd ed. Cold Spring Harbor Laboratory Press: Cold Spring Harbor, NY.
6. Brudno M, Chapman M, Göttgens B, Batzoglou S, Morgenstern B. (2003) Fast and sensitive multiple alignment of large genomic sequences *BMC Bioinformatics* 4:66
7. Iain M Wallace et al, Evaluation of Iterative Algorithms for Multiple Sequence Alignment, Bioinformatics Oxford Journal, Vol 21 No 8 2005.
8. Higgins DG, Sharp PM: Clustal: a package for performing multiple sequence alignment on a microcomputer. Gene 73, 237-244. 1988.
9. Thompson JD, Higgins DG, Gibson TJ. CLUSTAL W: improving the sensitivity of progressive multiple sequence alignment through sequence weighting, positions-specific gap penalties and weight matrix choice. Nucleic Acids Res 22:4673-4680. 1994.
10. http://align.genome.jp/
11. http://www.ebi.ac.uk/clustalw/
12. http://www.ch.embnet.org/software/ClustalW.html

13. Notredame C, Higgins DG, Heringa J: T-Coffee: a novel method for fast and accurate multiple sequence alignment. J Mol Biol, 302:205-217. 2000.
14. Notredame C, Higgins DG. SAGA: Sequence Alignment by Genetic Algorithm. Nucleic Acids Res 24(8):1515-24. 1996.
15. Notredame C, O'Brien EA, Higgins DG. RAGA: RNA Sequence Alignment by Genetic Algorithm. Nucleic Acids Res 25(22):4570-80. 1997.
16. Kim J, Pramanik S, Chung MJ. Multiple Sequence Alignment using Simulated Annealing. Comput Appl Biosci 10(4):419-26. 1994.
17. Mohd. Faizal Omar, Rosalina Abdul Salam, Nuraini Abdul Rashid, Rosni Abdullah. Multiple Sequence Alignment Using Genetic Algorithm and Simulated Annealing. ieeexplore.ieee.org/iel5/9145/29024/01307828.pdf. IEEE. 2004.
18. M. F. Omar, R. A. Salam, R. Abdullah, N. A. Rashid. Multiple Sequence Alignment Using Optimization Algorithms. INTERNATIONAL JOURNAL OF COMPUTATIONAL INTELLIGENCE 1(1): 87-95.2004.
19 M. Gen, R. Cheng. Genetic Algorithms and Engineering Optimization. John Wiley & Sons: Canada. 2000.
20 S. Kirkpatrick, C. D. J. Gellat, and M. P. Vecchi. Optimization by Simulated Annealing, Science, 220 pp. 671-680. 1983.
21 Qi-wen Dong, Lei Lin, Xiao-Long Wang and Ming-Hui Li. Contact-based Simulated Annealing Protein Sequence Alignment Method. Engineering in Medicine and Biology Society, 2798- 2801. Sept. 2005.